\documentclass[usenatbib,usenatbib]{mnras}
\usepackage{newtxtext,newtxmath}
\usepackage{todonotes}
\usepackage{times}
\usepackage{newtxtext,newtxmath}

\usepackage{amsmath, bm}
\usepackage{graphicx} 
\usepackage{bmpsize}
\usepackage{epstopdf}
\usepackage{dblfloatfix} 
\graphicspath{{./images/}}
\usepackage[normalem]{ulem}
\usepackage[T1]{fontenc}
\usepackage{makecell}
\usepackage{multirow}

\newcommand{\HI}{\rm H{\sc i }}

\newcommand{\TB}{\delta T_{\rm b}}
\newcommand{\MSUN}{{\rm M}_{\odot}}

\newcommand{\XHI}{x_{\rm HI}}
\newcommand{\XHII}{x_{\rm HII}}
\newcommand{\TS}{T_{\rm S}}
\newcommand{\TK}{T_{\rm K}}
\newcommand{\TCMB}{T_{\gamma}}

\newcommand{\lya}{\rm {Ly{\alpha}}}
\newcommand{\OmegaB}{\Omega_{\rm B}}
\newcommand{\Omegam}{\Omega_{\rm m}}

\title[IGM during the Cosmic Dawn]{The morphology of the redshifted 21-cm signal from the Cosmic Dawn}

\author[Ghara et al.]{Raghunath Ghara$^{1,2,3}$\thanks{E-mail: \href{mailto:ghara.raghunath@gmail.com}{ghara.raghunath@gmail.com}}, Satadru Bag$^{4,5}$, Saleem Zaroubi$^{1,5,6}$, Suman Majumdar$^{7}$\\
$^1$Astrophysics Research Center of the Open University (ARCO), The Open University of Israel, 1 University Road, Ra'anana 4353701, Israel.\\
$^2$Haverford College, 370 Lancaster Ave, Haverford PA, 19041, USA\\
$^3$Center for Particle Cosmology, Department of Physics and Astronomy, University of Pennsylvania, Philadelphia, PA 19104, USA\\
$^4$ Department of Physics, TUM School of Natural Sciences, Technical University of Munich,  James-Franck-Straße 1, 85748 Garching, Germany\\
$^5$ Max-Planck-Institut fur Astrophysik, Karl-Schwarzschild-Str. 1,  85748 Garching, Germany \\
$^6$Kapteyn Astronomical Institute, University of Groningen, PO Box 800, 9700AV Groningen, The Netherlands.\\
$^7$Department of Astronomy, Astrophysics \& Space Engineering, Indian Institute of Technology Indore, Indore 453552, India}

\date{Accepted XXX. Received YYY; in original form ZZZ}

\pubyear{}

\begin{document}
\label{firstpage}
\pagerange{\pageref{firstpage}--\pageref{lastpage}}
\maketitle
\begin{abstract}
The spatial fluctuations in the tomographic maps of the redshifted 21-cm signal from the Cosmic Dawn (CD) crucially depend on the size and distribution of the regions with gas temperatures larger than the radio background temperature. In this article, we study the morphological characteristics of such emission regions and their absorption counterparts using the shape diagnostic tool {\sc surfgen2}. Using simulated CD brightness temperature cubes of the 21-cm signal, we find that the emission regions percolate at stages with the filling factor of the emission regions $\mathrm{FF}_{\rm emi}\gtrsim 0.15$. Percolation of the absorption regions occurs for $\mathrm{FF}_{\rm abs}\gtrsim 0.05$. The largest emission and absorption regions are topologically complex and highly filamentary for most parts of the CD. The number density of these regions as a function of the volume shows the power-law nature with the power-law indexes $\approx -2$ and $-1.6$ for the emission and absorption regions, respectively. Overall, the planarity, filamentarity and genus increase with the increase of the volume of both emission and absorption regions.

\end{abstract}



\section{INTRODUCTION}
\label{sec:intro}
Our Universe became neutral during the epoch of recombination about four hundred thousand years after the Big Bang. At a later stage, radiation emitted from the first stars, galaxies, Quasars, High-mass X-ray binaries (HMXBs), etc. heat up and ionize the cold and neutral gas in the inter-galactic medium (IGM). Theoretical studies such as \citet[][]{Pritchard07,2011MNRAS.411..955M, ghara15c, 2019MNRAS.487.2785I, Ross2019} suggest that the first X-ray sources such as HMXBs, and mini-Quasars heat up the gas in the IGM during a period earlier than the stage when the IGM became highly ionized. This period is known as the Cosmic Dawn (CD) of our Universe. The CD was succeeded by a period, known as the Epoch of Reionization (EoR), in which the IGM's primordial neutral gas became ionized by the UV radiation from the sources. The physical processes of these eras have major roles in shaping the current state of our Universes through various feedback mechanisms. Unfortunately, these epochs remain among the least understood epochs of Cosmic history \citep[e.g.][]{Fan06b, 2018Natur.553..473B, Planck2018, Mitra15}. Many details about the timing, feedback mechanisms, morphology of the ionized and heated regions, etc. are still poorly known. 

Observation of the redshifted 21-cm line radiation emitted from the IGM's neutral hydrogen (\HI) atoms is the most promising probe of the physical processes during the CD and EoR \citep[see e.g.,][]{madau1997, Furlanetto2006, 2013ASSL..396...45Z, 2022JApA...43..104M, Shaw_review, 2023JApA...44...10B}. Many of the world's largest radio telescopes have dedicated their valuable resources to measuring the \HI 21-cm signal from these periods. One type of such observations uses radio interferometers such as  the Low-Frequency Array (LOFAR)\footnote{\url{http://www.lofar.org/}} \citep{2020MNRAS.493.1662M}, the Precision Array for Probing the Epoch of Reionization (PAPER)\footnote{\url{http://eor.berkeley.edu/}} \citep{ 2019ApJ...883..133K}, the Murchison Widefield Array (MWA)\footnote{\url{http://www.mwatelescope.org/}} \citep[e.g.][]{Wayth2018mwa} and the Hydrogen Epoch of Reionization Array (HERA)\footnote{\url{https://reionization.org/}} \citep{2017PASP..129d5001D}, the New Extension in Nançay Upgrading LOFAR (NenuFAR)\footnote{ \url{https://nenufar.obs-nancay.fr/en/homepage-en/}}\citep{refId0}, the Amsterdam ASTRON Radio Transients Facility And Analysis Center (AARTFAAC)\citep{2022A&A...662A..97G}. Due to limited sensitivity, the ongoing radio interferometer-based 21-cm signal observations of the CD and EoR aim to measure the time evolution of the spatial fluctuations of the signal at different scales \citep[e.g.,][]{2020MNRAS.499.4158G, 2020MNRAS.493.1662M, 2020MNRAS.493.4711T, Abdurashidova_2023, refId0}. The upcoming Square Kilometre Array (SKA)\footnote{\url{http://www.skatelescope.org/}} will have a higher sensitivity and will further produce tomographic images of the distribution of the \HI signal on the sky \citep{2015aska.confE..10M, ghara16}. The second type of radio observations such as EDGES \citep{2010Natur.468..796B}, EDGES2 \citep{monsalve2017, EDGES2018}, SARAS \citep{2015ApJ...801..138P}, SARAS2 \citep{singh2017}, LEDA \citep{price2018} use single radio antenna and aim to measure the redshift evolution of the sky averaged \HI 21-cm signal from these epochs.

The redshifted \HI 21-cm signal from the CD and EoR is enriched with information about the first sources of our Universe. The measurements of the redshifted \HI signal from these epochs are often used to infer the properties of the sources formed during these epochs \citep[e.g.,][]{2020MNRAS.493.4728G, 2020arXiv200603203G, 2020MNRAS.498.4178M, 2022ApJ...924...51A}, nature of dark matter \citep[e.g.,][]{2018Natur.555...71B, 2018PhRvL.121a1101F, 2018Natur.557..684M, 2019JCAP...04..051N,  2020MNRAS.492..634G, 2022JCAP...03..055G}, etc. Furthermore, the same observations can be used to infer the properties of the IGM such as the thermal and ionization states and morphology of the 21-cm signal maps \citep[e.g.,][]{2020MNRAS.493.4728G, 2020MNRAS.496..739G, 2021MNRAS.503.4551G}. The EoR 21-cm signal is expected to be enriched with information about the ionized regions while the CD 21-cm signal is expected to carry information about the characteristics of the heated/emission regions. The heated regions around the early X-ray emitting sources such as the HMXBs, mini-Quasars are expected to appear as emissions in the Cosmic microwave background (CMB) subtracted 21-cm signal maps while the cold regions will appear as absorption signal \citep[see e.g.,][]{ghara15c, ghara16}.

The radio interferometer-based observations probe multi-scale properties of the signal and thus, are expected to contain more information about the states of the IGM. The fluctuation in the observed 21-cm signal by the radio interferometers depends on the morphology and distribution of the ionized/emission regions. Thus it is important to study the morphology of the ionized/emission regions of the signal to understand the connection between the measured statistical quantities of the signal such as the power spectrum or the sky-averaged signal. Recently several studies using methods such as {\sc granulometry} method \citep{2017MNRAS.471.1936K}, Mean-free path \citep{mesinger07}, image segmentation method \citep{2018arXiv180106550G},  percolation analyses \citep{Iliev2006, Iliev2014, Furlanetto2016} and the Minkowski functionals \citep{Friedrich2011, yoshiura:2017, Kapahtia2021}, etc. have made a significant effort in understanding the morphology of the ionized and neutral regions during the EoR. These studies indicate that the characteristics of the ionized/neutral regions e.g., the probability density functions (PDF) of the bubble size distribution can be used to distinguish different reionization scenarios and thus to constrain the EoR source parameters, etc.

\begin{figure*}
\begin{center}
\includegraphics[scale=0.54]{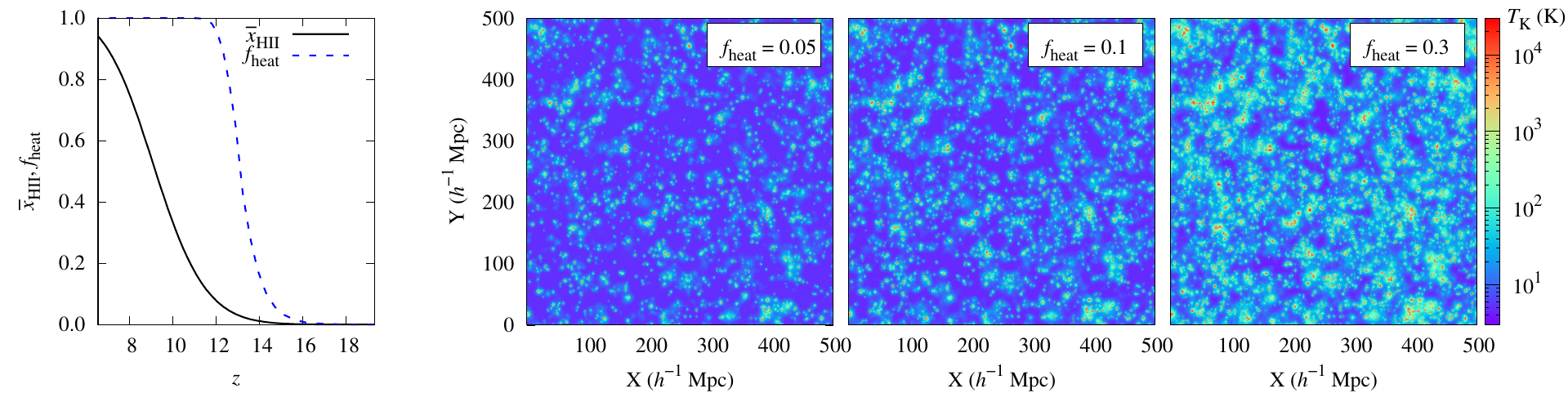}
    \caption{Ionization and heating history of the simulated Cosmic Dawn and reionization scenario. The left panel shows the evolution of the volume fraction of the ionized region (i.e., volume averaged ionization fraction $\bar{x}_{\rm HII}$) and heated regions (i.e., heated fraction $f_{\rm heat}$) of our simulated Cosmic Dawn (CD) and Epoch of Reionization (EoR) as functions of redshift. We define a `heated region' as a region with a gas temperature larger than the CMB temperature. The left to right two-dimensional maps show the gas temperature at three different stages of the CD with heated fractions 0.05, 0.1 and 0.3, respectively. The corresponding redshifts are 14.7, 14.3 and 13.5, respectively. The colour bar shows the temperature in Kelvin.}
   \label{Fig.tk}
\end{center}
\end{figure*} 
  
In this work, we focus on the morphological analysis of the \HI 21-cm signal from the CD. Similar to the ionized/neutral regions which are the focus of morphology studies of the 21-cm signal from the EoR  \citep[e.g.,][]{mesinger07, 2017MNRAS.471.1936K, 2018MNRAS.477.1984B, Kapahtia2021}, we consider the emission/absorption regions and study the morphology and distribution of these regions. Our study is based on simulated 21-cm signal maps and {\sc surfgen2} algorithm \citep{2019MNRAS.485.2235B, Bag_sdss} which assesses the geometry, morphology and topology of a field above/below certain thresholds. This algorithm models the surface of a region through the {\em Marchingcube 33} triangulation \citep{marcube,mar33}. The accuracy in measuring the Minkowski functionals and Shapefinders is much better compared to the other existing methods, such as the cell counting-based Crofton's formula \citep{Crofton_1968}. Previously, \citet[][]{2018MNRAS.477.1984B, 2019MNRAS.485.2235B, 2022arXiv220203701P, Dasgupta20233} studied the morphology of the ionized regions during the EoR using the same algorithm. These studies found that the time evolution of the Largest Cluster Statistics (LCS) of the ionized regions is a robust means to study the percolation of the ionized regions and can be used to distinguish between different reionization scenarios.  Here, we study the evolution of the emission regions LCS and the percolation state for the emission regions of the 21-cm signal during the CD. In fact, we would like to point out that this paper is the first effort to use LCS for the morphological analysis of the CD 21-cm signal, all previous work with LCS has been limited to 21-cm signal coming from the EoR.

This paper is structured in the following way. We first describe the simulation used in this study in Section \ref{sec:method}, then we discuss the methodology to characterise the emission regions in Section \ref{sec.mor}.  Section \ref{sec:results} shows the results from this study. We discuss and conclude the findings of this study in Section \ref{sec:con}. This study uses cosmological parameters   $\Omegam=0.27$, $\Omega_\Lambda=0.73$, $\OmegaB=0.044$, $h=0.7$  \citep[Wilkinson ~Microwave ~Anisotropy ~Probe (WMAP);][]{2013ApJS..208...19H} which are the same as used in the N-body simulation used in this study.

\begin{figure*}
\begin{center}
\includegraphics[scale=0.54]{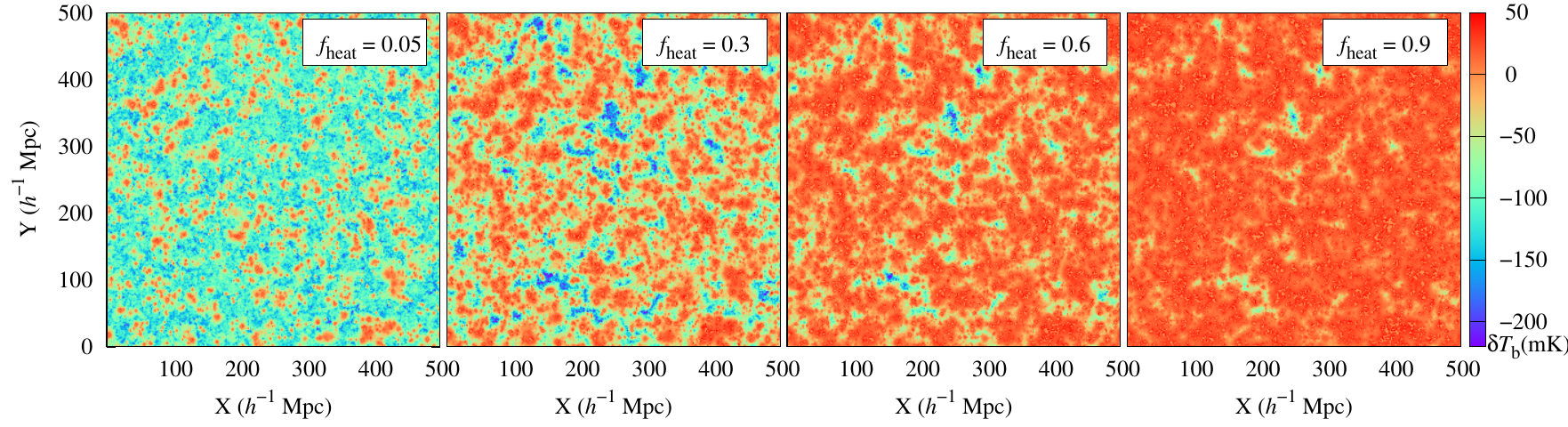}
    \caption{Slices through the $\TB$ cubes at different stages of the Cosmic Dawn. From left to right panels correspond to redshifts 14.7, 13.5, 12.9 and 12.3 with heated fractions 0.05,  0.3, 0.6, and 0.9, respectively. These maps correspond to $\TCMB=T_{\rm CMB}$. }
   \label{Fig.tbmap}
\end{center}
\end{figure*}

\section{Simulation}
\label{sec:method}
In this section, we briefly present the simulation that leads towards 21-cm signal brightness temperature maps used for the morphological study. Here, we use {\sc grizzly}  \citep{ghara15a, ghara18} code to simulate the redshifted 21-cm signal from the CD and EoR. The algorithm is based on one-dimensional radiative transfer (RT) and is an independent implementation of the previous 1D RT-based code {\sc bears}  \citep{Thom09, 2018NewA...64....9K}.

The differential brightness temperature ($\TB$) of the redshifted \HI 21-cm signal emitted from a region with angular position $\mathbfit{x}$ and redshift $z$  can be expressed as \citep[see e.g.,][]{madau1997}
\begin{equation}
    \TB(\mathbfit{x}, z)=\frac{\TS(\mathbfit{x}, z)-\TCMB(z)}{1+z}\left[1-\exp\{-\tau_{\mathrm{b}}(\mathbfit{x}, z)\}\right], 
    \label{eq:brightnessT}
\end{equation}
where the \HI 21-cm signal optical depth  $\tau_{\mathrm{b}}$ can be denoted as \citep{bharadwaj05, 2022MNRAS.509..945D},
\begin{eqnarray}
\tau_{\mathrm{b}}(\mathbfit{x}, z) &=& \frac{4 \, {\rm mK}}{\TS(\mathbfit{x}, z)} x_{\rm HI} (\mathbfit{x}, z) [1+\delta_{\rm B}(\mathbfit{x}, z)] \nonumber \\ 
&\times& \left( \frac{\OmegaB h^2}{0.02}\right) \left(\frac{0.7}{h}\right)
\frac{H_0}{H(z)}(1+z)^3.
\label{eq:tau}
\end{eqnarray}
Here, $\TCMB$ is the brightness temperature of the radio background. In the absence of exotic sources of radio emission \citep[e.g., see][]{2020MNRAS.498.4178M, 2022JCAP...03..055G},   $\TCMB$ is expected to be the CMB brightness temperature  $\TCMB(z)$ = 2.73 $\times (1+z)$ K. The quantities $\XHI$ and $\delta_{\rm B}$ are the neutral fraction and the density contrast of hydrogen in that region, respectively. The quantity $\TS$ stands for the spin temperature of the neutral hydrogen which quantifies the relative population of hydrogen in the two hyperfine states of the hydrogen atom's ground state.

 The {\sc grizzly} algorithm uses dark matter halo catalogues and density and velocity fields on grids as derived from N-body simulations. Given these inputs, {\sc grizzly} generates $\TB$ maps for a source model.  Here, we use the outcome of a dark-matter-only N-body simulation from the results of the PRACE\footnote{Partnership for Advanced Computing in Europe: \url{http://www.prace-ri.eu/}} project PRACE4LOFAR. The simulation was performed using {\sc cubep$^3$m} \citep{Harnois12} code with 6912$^3$ particles in a box of volume $(500 ~h^{-1}~\rm Mpc)^3$ which set the particle mass resolution to $4.05\times 10^7 ~\MSUN$. Later, the density and velocity fields are smoothed into grids. We use $300^3$ grids in this paper. The dark matter halo catalogues were generated on the fly with a spherical overdensity halo finder \citep{Watson2013TheAges}. These contain all halos with masses larger than  $10^9$~M$_\odot$ which is $\approx 25$ times the dark matter particles resolution. The simulation saved the gridded fields and halo catalogues at different redshifts while the time gap between two successive redshifts is 11.4 mega-years.  Details of the N-body simulation can be found in  \citet[][]{Dixon2016TheReionization, 2019JCAP...02..058G}. Here we use 59 redshift snapshots between $z\sim 20-6.5$ to simulate the CD and EoR for the following source model.   

The astrophysical sources dependence in Equation \ref{eq:brightnessT} comes through $\XHI$ and $\TS$. Both these quantities depend on the emissivity of UV, X-ray, radio and $\lya$ photons, as well as the population number density of the sources \citep[see e.g.,][]{Ross2019, 2020MNRAS.499.5993R}. The source model adopted in this study assumes that each of the dark matter halos on the halo catalogues emits UV, X-ray and $\lya$ photons. The emission rates of these photons are proportional to the stellar mass ($M_\star$) of the sources which we assume to be linearly related to the dark matter halo mass  $M_{\rm halo}$, i.e.,  $M_\star=f_\star \left(\frac{\OmegaB}{\Omegam}\right) M_{\rm halo}$. We fix the star formation efficiency  $f_\star=0.02$ in this study. The emission rate of the UV photons per stellar mass $\dot N_i$ is controlled by the Ionization efficiency parameter  $\zeta$ as $\dot N_{\mathrm{i}}=\zeta\times 2.85\times 10^{45}  ~{\rm s^{-1}} ~\MSUN^{-1}$. The cosmic heating and reionization depends also on the minimum mass of dark matter halos ($M_{\rm min}$) that contain radiating sources. Here, we have assumed that all dark matter halos with masses larger than $M_{\rm min}=10^9~\MSUN$ host sources of UV, X-rays and $\lya$ photons. We use $\zeta=0.1$ for which the EoR ends around $z\sim 6.5$\footnote{Note that our simulated reionization scenario ends slightly earlier compared to the time as constrained by some observations such as high-redshift quasar absorption spectra \citep[see e.g.,][]{Fan06b, 2018Natur.553..473B}. However, the end redshift of reionization is still debated and it is unlikely that this will change the conclusion of the work as our study focuses on the high-redshift Cosmic Dawn era. Therefore, we choose a reionization simulation from our earlier works such as \citet{2023MNRAS.522.2188S}.}.   The X-ray spectral energy distribution of the source is assumed to be a power-law with energy $E$ as $I_{\mathrm{X}}(E) \propto E^{-\alpha}$ with spectral index $\alpha=1.2$. The emission rate of the X-ray photons per stellar mass  $\dot N_{\mathrm{X}} = f_{\mathrm{X}} \times  10^{42} ~\rm s^{-1} ~\MSUN^{-1}$ where $f_{\mathrm{X}}$ is the X-ray heating efficiency parameter. The X-ray band spans from 100 eV to 10 keV. We choose $f_{\mathrm{X}}=100$ for the heating model considered in this study. The details of the spectral energy distribution and source model can be found in \citet[][]{ghara15a,2019MNRAS.487.2785I}. These references also provide the details of the method to simulate the coeval cubes of neutral fraction $\XHI$ and $\TK$ at different redshifts.

The solid line of the left panel of Figure \ref{Fig.tk} shows the redshift evolution of the ionization fraction while the dashed curve shows the evolution of the heated fraction $f_{\rm heat}$ i.e., the volume fraction of regions with gas temperature larger than the CMB brightness temperature.  The Figure shows that the IGM is significantly heated by redshift $\sim 13$ while the ionization fraction remains less than 0.1. Figure \ref{Fig.tk} also shows slices of the gas temperature distribution at three different stages of heating. From left to right $\TK$ maps represent redshifts  14.7, 14.3 and 13.5 with $f_{\rm heat}$ 0.05, 0.1 and 0.3, respectively. The slices show that significant overlap between heated regions already started as early as a stage with redshift 14.3.  The $\lya$ photon flux fields at different redshifts are generated assuming a $1/R^2$ fall of the number density of $\lya$ photons with radial distance $R$ from the source. These $\lya$ photon flux fields are later used to generate the $\TS$ fields.  These $\XHI$, $\TS$ cubes and the density fields are then used to generate the $\TB$ maps using Equation \ref{eq:brightnessT}. Finally, we incorporate the effect of the peculiar velocities of the gas or so-called redshift-space distortion using a cell movement method \citep[see e.g.,][]{mao12, 2021MNRAS.506.3717R}. Figure \ref{Fig.tbmap} shows the $\TB$ slices at four different stages of the CD. The size of the heated/emission regions (with $\TB\geq 0$) increases as heating progresses. The panels also show that, above $f_{\rm heat}\gtrsim 0.5$, the distribution of the absorption regions (with $\TB<0$) is more relevant compared to the emission regions.

\begin{figure*}
\begin{center}
\includegraphics[scale=0.54]{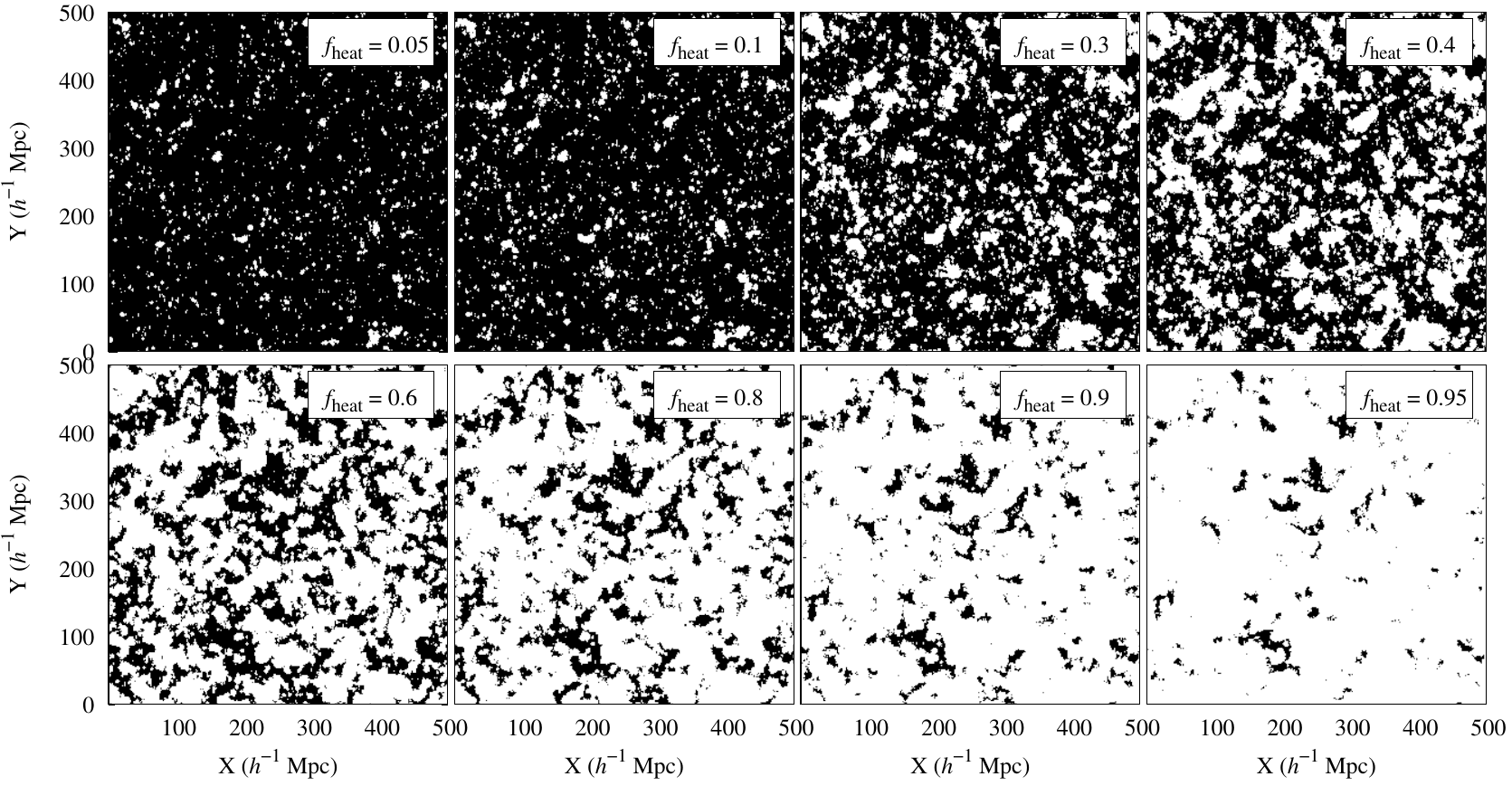}
    \caption{Slices through the binary cubes ${\TB}^{\rm bin}$ of the \HI 21-cm signal brightness temperature at different stages of the CD. From left-top to right-bottom panels corresponds to redshifts 14.7, 14.3, 13.5, 13.2, 12.9, 12.6 12.3 and 12.0 with heated fractions 0.05, 0.1, 0.3, 0.4, 0.6, 0.8, 0.9 and 0.95, respectively. The binary cubes ${\TB}^{\rm bin}(\mathbfit{x}, z)$ assume ${\TB}^{\rm bin}(\mathbfit{x}, z)=1$ if $\TB(\mathbfit{x}, z) \geq 0$ ( white regions), and zero otherwise (black regions). These maps correspond to $\TCMB=T_{\rm CMB}$. }
   \label{Fig.bin}
\end{center}
\end{figure*}

\section{Morphology analysis}
\label{sec.mor}
There exists no unique way to characterize the morphology and size distribution of a complex spatial structure such as the distribution of $\TB$ emission regions in the IGM. Here we use {\sc surfgen2} algorithm \citep{2019MNRAS.485.2235B, Bag_sdss} to assess the morphology of individual heated regions in the 3D coeval $\TB$ boxes generated using {\sc grizzly}. Given a coeval cube of the $\TB$, we first generate a binary field ${\TB}^{\rm bin}(\mathbfit{x}, z)$ where ${\TB}^{\rm bin}(\mathbfit{x}, z)=1$ if $\TB(\mathbfit{x}, z) \geq 0$, and zero otherwise. These binary maps are then used in {\sc surfgen2} for the morphological analysis of the emission regions. The code implements periodic boundary conditions on the simulated $\TB^{\rm bin}(\mathbfit{x}, z)$ boxes and uses the friends-of-friends (FoF) algorithm to identify individual emission and absorption regions. Figure \ref{Fig.bin} shows slices of such binary fields at different stages of our simulated CD as described in the previous section. The white regions in the slices represent emission regions for $\TCMB=T_{\rm CMB}$. For $f_{\rm heat}>0.5$, the size distribution of the absorption regions is more meaningful than the emission regions. This is due to the inside-out nature of the heating around the X-ray sources. Clearly, the individual emission regions grow in size over time and overlap to create a large region that dominates the total emission segment volume, while the morphology of the absorption regions towards the end of the CD appears different in nature. Visually the morphology of these binary emission regions is similar to the morphology of ionized regions during an inside-out reionization scenario \citep[see e.g.,][]{2018MNRAS.477.1984B, 2022arXiv220203701P}. 

The calculation of the Minkowski functionals and thereafter the Shapefinders of a connected (emission/absorption) region in {\sc surfgen2} is based on modelling the surface of that region through the advanced {\em Marching Cube 33} triangulation scheme \citep{marcube,mar33}. We refer the readers to \citet{Sheth:2002rf,2019MNRAS.485.2235B} papers for the details of the {\sc surfgen2} algorithm. The algorithm first determines the four basic Minkowski functionals \citep{mecke} that well describe the morphology of an individual region. These are
\begin{enumerate}
 \item volume: $V$,
 \item surface area: $S$,
 \item integrated mean curvature (IMC):
 \begin{equation}
  \mathrm{IMC}=\frac{1}{2} \oint_S \left(\kappa_1+\kappa_2\right) dS \;,
 \end{equation}
  \item integrated Gaussian curvature or Euler characteristic: 
  \begin{equation}
  \chi=\frac{1}{2\upi} \oint_S (\kappa_1 \kappa_2) dS\;.
  \end{equation}
\end{enumerate}
Here $\kappa_1$ and $\kappa_2$ represent the two principle curvatures at the surface element $dS$ on the surface $S$. The integrated Gaussian curvature $\chi$ is related to the genus ($G$) of a closed surface as  $G=1-\chi/2$. Physically $ G\equiv {\rm (no.~ of~ tunnels)}-{\rm (no.~ of~ isolated~ surfaces)}+1$. Note that while $V$ and $S$ of the emission and absorption regions will always be positive, $\mathrm{IMC}$ and $G$ might also be negative in certain cases.

To measure the shape of a region directly, we calculate the following three ratios of the Minkowski functionals,
\begin{equation}
T=3V/S \;, ~~\\  B=S/|\text{IMC}]| \;, ~~\\ L=|\text{IMC}|/(4\upi).
\label{eq.shape}
\end{equation}
Note that each of $T$, $B$ and $L$ are of the dimension of length and are spherically normalized, i.e. $V=(4\upi/3) TBL $. These ratios were first introduced in \citet{Sahni:1998cr} as `Shapefinders' since they assess the extent of a region in three dimensions. Here, $L, ~B$ and $T$ represent the length, breadth and thickness of a region, respectively. A sphere of radius $R$ will have $T=B=L=R$. These quantities should satisfy natural ordering $L\geq B \geq T$. However, this ordering might not be maintained naturally in some rare cases. In those cases, we define the largest value among the three in Equation \ref{eq.shape} as $L$ and the smallest value as $T$, to maintain the order $L\geq B \geq T$.

Further, the planarity ($P$) and filamentarity ($F$) of a region are defined as the following dimensionless quantities based on the Shapefinders \citep[see][]{Sahni:1998cr},  
\begin{equation}\label{eq:PF}
 P=\frac{B-T}{B+T}\;, ~~\\
 F=\frac{L-B}{L+B}.
\end{equation}
As their names suggest, they measure the degree of planarity and filamentarity of a region. By definition, $P$ and $F$ values are bounded by 0 and 1. Note that, for a spherical object, $T=B=L$ leads to $P=0=F$. For a planar object, such as a sheet, $L\sim B \gg T$ results in $P \gg F$. On the other hand, $T \sim B \ll L$ for a filament and thus $F \gg P$.

In this work, we employ the above-mentioned Minkowski functionals and Shapefinders to statistically study the morphology of the emission and absorption regions of our simulated CD 21-cm signal. In addition, we consider the morphological properties of the largest emission and absorption regions which provides crucial information about the percolation state of these regions.

\section{Results}
\label{sec:results}

\begin{figure}
\begin{center}
\includegraphics[scale=0.61]{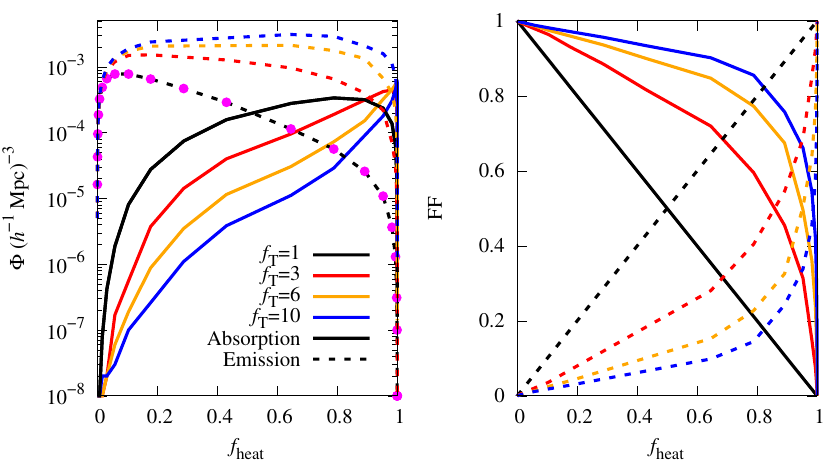}
    \caption{Evolution of the number density of emission and absorption regions (left panel) and their filling factors $\mathrm{FF}$ (right panel) as functions of the heated fraction, i.e., the volume fraction of regions with gas temperature larger than $T_{\rm CMB}$. Different types of lines represent different threshold parameter ($f_\mathrm{T}$) values for the radio background temperature as chosen in this study. The magenta circular dots mark different stages of the CD where the morphology of the emission regions has been estimated. The emission and absorption regions at different stages of the CD are determined using {\sc surfgen2} algorithm. }
   \label{Fig.stat}
\end{center}
\end{figure}

In this paper, we explore the morphological distribution of the emission and absorption regions of the 21-cm signal during the CD. We consider simulated $\TB$ coeval maps at different redshifts generated from {\sc grizzly} code (see section \ref{sec:method}). Unlike the ionization maps as considered previously in studies like \citet[][]{2018MNRAS.477.1984B, 2022arXiv220203701P}, the fall of the signal amplitude at the boundary of the emission regions where $\TB=0$ is not very sharp. Furthermore, the nature of the signal, either for emission or absorption, depends crucially on the radio background temperature $\TCMB$ which is uncertain at the frequencies of our interest  \citep[e.g.,][]{2011ApJ...734....5F}.  Thus,  we choose a number of different thresholds $\delta T_{\rm b,thre}$ on the $\TB(\mathbfit{x}, z)$ maps to study the morphology of the regions with $\TB$ larger than $\delta T_{\rm b,thre}$ which we refer to as emission segment. In addition, we also consider the complementary regions, i.e. the absorption regions with $\TB< \delta T_{\rm b,thre}$. We have seen from Figures \ref{Fig.tbmap} and \ref{Fig.bin} that the distribution of these absorption regions is more relevant for $f_{\rm heat}\gtrsim 0.5$. Different $\delta T_{\rm b,thre}$ values also represent different strengths of radio background at 1420 MHz at the redshift of our interest. We parameterize the  $\delta T_{\rm b,thre}$ in terms of radio background temperature $T_{\gamma}=f_\mathrm{T} \times 2.725(1+z)$ K with the threshold parameter $f_\mathrm{T}$. Note that, similar to the majority of the 21-cm signal study, $f_{\mathrm T}=1$ is the default threshold for our study and it will be the same if not mentioned otherwise.

The number density and size of the emission regions vary with the radio background temperature $T_{\gamma}$ and thus also with threshold parameter $f_\mathrm{T}$. The left panel of Figure \ref{Fig.stat} shows the evolution of the number density ($\Phi$) of isolated emission (dashed curves) and absorption (solid curves) regions with $f_{\rm heat}$ for four different $f_\mathrm{T}$ values. The magenta circular points in the left panel of Figure \ref{Fig.stat} mark different CD stages where the morphology of the emission and absorption regions have been estimated. $\Phi$ of the emission regions first increases with $f_{\rm heat}$ until $f_{\rm heat}$ reaches $\sim 0.1$ due to the formation of new heated regions. For $f_{\rm heat}\gtrsim 0.1$, $\Phi$ of the emission regions gradually falls before sharply falling to 1 when $f_{\rm heat}\rightarrow 1$.  This happens due to overlaps between individual emission regions.  Both the maximum value of emission $\Phi$ and the corresponding $f_{\rm heat}$ (at which the maximum occurs) remain smaller for a smaller value of $f_\mathrm{T}$. This is expected as by increasing $f_\mathrm{T}$, a large emission region at a later stage of heating can break into multiple smaller emission regions. The fall of $\Phi$ for emission regions is faster for a smaller value of $f_\mathrm{T}$. In fact, $\Phi$ changes very little in the range $0.1\lesssim f_{\rm heat}\lesssim 0.9$ for $f_\mathrm{T}=10$. This implies that the overlap between the emissions regions happens at a later stage for a larger $f_\mathrm{T}$ value. 

 \begin{figure}
\begin{center}
\includegraphics[scale=0.68]{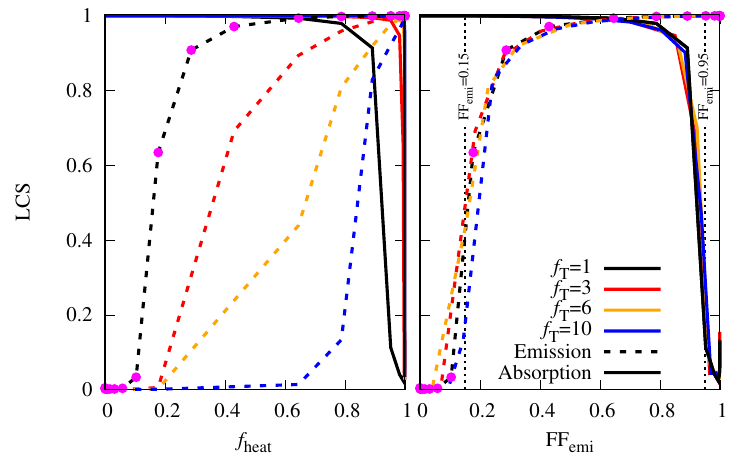}
    \caption{Evolution of the Largest cluster statistics (LCS) as a function of the heated fraction (left panel) and filling factor (right panel) of the emission regions. The dashed curves represent emission regions while the solid curves represent the absorption regions of the 21-cm signal $\TB$ coeval cube. The magenta circular points mark different stages of CD when the morphology of the emission and absorption regions has been estimated. The largest emission and absorption regions at different redshifts of the CD are determined using {\sc surfgen2} algorithm. }
   \label{Fig.statlcs}
\end{center}
\end{figure}

The number density of absorption regions, on the other hand, gradually increases with $f_{\rm heat}$ since larger absorption regions break into smaller ones as heating progresses. Contrary to the $\Phi$ of the emission regions, the $\Phi$ of the absorption regions (solid curves of Figure \ref{Fig.stat}) decreases with the increase of $f_\mathrm{T}$. However, $\Phi$ of absorption regions sharply drops to zero at the end of the heating of the IGM, as expected.  

The filling factor (which is defined as the volume fraction of emission or absorption segment), $\mathrm{FF}$, of emission or absorption regions also depends on $f_\mathrm{T}$. The right panel of Figure \ref{Fig.stat} shows how $\mathrm{FF}$s of the emission (dashed) and absorption (solid) regions evolve with $f_{\rm heat}$ for different $f_\mathrm{T}$ values. As expected, $\mathrm{FF}$ of the emission regions ($\mathrm{FF}_{\rm emi}$) is roughly the same as the $f_{\rm heat}$ for $f_\mathrm{T}=1$. For a certain $f_{\rm heat}$, $\mathrm{FF}_{\rm emi}$ decreases for a larger value of $f_\mathrm{T}$. On contrary to $\mathrm{FF}_{\rm emi}$, $\mathrm{FF}$ of the absorption regions ($\mathrm{FF}_{\rm abs}$) decreases as heating progresses since $\mathrm{FF}_{\rm abs}=1-\mathrm{FF}_{\rm emi}$.

\subsection{Largest cluster statistics}
The `largest cluster statistics' (LCS) for a segment (emission or absorption) is defined as the volume fraction enclosed by the largest region of that segment,
\begin{equation}
    \text{LCS}=\frac{\text{volume of the largest emission/absorption region}}{\text{total volume of all the emission/absorption regions}}\;.
\end{equation}
 It has been established that the LCS carries robust information about the percolation process in different fields at cosmological scales, e.g. the large-scale matter distribution in the Universe \citep{Shandarin1983, Klypin1993, Yess1996} and the neutral hydrogen field during the EoR \citep{Iliev2006, Iliev2014, Furlanetto2016, 2018MNRAS.477.1984B}. For the ionization bubbles, the percolation happens early with volume averaged ionization fraction $\overline{x}_{\rm HII}\sim 0.1$ \citep[see e.g.,][]{Furlanetto2016, 2018MNRAS.477.1984B}. Here, we study the characteristics of the largest emission and absorption regions in the IGM at different stages of CD and the percolation transitions in the respective segment.

In the left panel of Figure \ref{Fig.statlcs}, we present the evolution of LCS of the emission and absorption regions as a function of $f_{\rm heat}$ for the four different choices of $f_\mathrm{T}$. At the initial stages of the heating (e.g., with $\mathrm{FF}_{\rm emi}\lesssim 0.1$), the emission regions are fairly isolated with nominal overlaps with each other. As heating progresses, many new emission regions appear around the X-ray emitting sources and the emission regions themselves grow in size resulting in a gradual increase in the LCS of the emission regions with $f_{\rm heat}$ (see the dashed curves in the left panel of Figure \ref{Fig.statlcs}). Soon the emission regions start to merge significantly. Due to the vigorous merging of smaller regions, an extremely large connected emission region that extends throughout the IGM appears abruptly at this stage. In our case, it stretches throughout the simulation box. We refer to this period as the `percolation transition' in the emission regions. The LCS of the emission regions increases sharply during the percolation transition since the largest region abruptly accumulates most of the emission segment volume. The critical $f_{\rm heat}$ at which percolation transition takes place depends on $f_\mathrm{T}$ values. However, percolation transition roughly occurs at similar values of the filling factor $\mathrm{FF}_{\rm emi}\sim 0.15$ (marked by the vertical dotted line in the right panel of Figure \ref{Fig.statlcs}) for all the threshold values chosen (see the dashed curves of the right panel of Figure \ref{Fig.statlcs}).

The LCS of the absorption regions, on the other hand, remained $\approx 1$ for a long time before sharply dropping to zero around $f_{\rm heat}\sim 1$. This suggests that the absorption segment in the IGM remains percolated for a large extent of the CD and only breaks into isolated absorption regions towards the end period of the CD. When plotted against $\mathrm{FF}$, the LCS of the absorption regions shows a sharp decrease around $\mathrm{FF}_{\rm emi}\sim 0.95$ for all the thresholds chosen.  The evolution of the LCS  of the emission regions with $\mathrm{FF}_{\rm emi}$ is almost identical for the different choices of $f_\mathrm{T}$.

\begin{figure*}
\begin{center}
\includegraphics[scale=0.56]{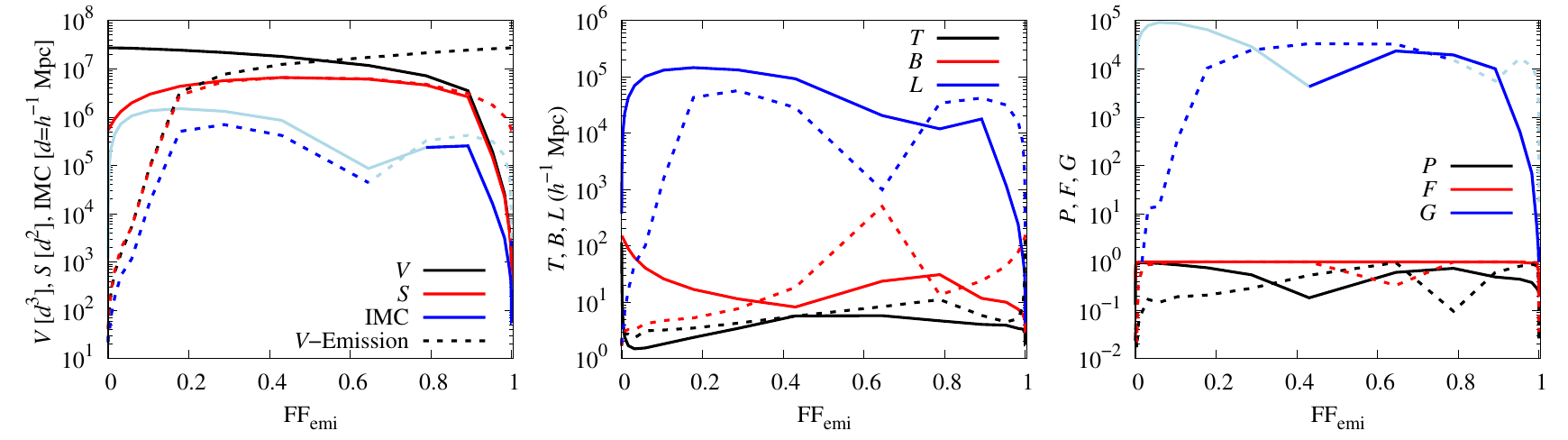}
    \caption{Evolution of different shape defining quantities of the largest region of the emission (dashed) and absorption (solid) segments in the IGM as functions of filling factor of the emission regions correspond to $f_\mathrm{T}=1$. The x-axis is equivalent to the heated fraction of the simulation. The left panel shows the volume ($V$), surface area ($S$) and integrated mean curvature  ($\mathrm{IMC}$) of the largest emission and absorption region. The middle panel shows the thickness ($T$), breadth ($B$) and length ($L$) of the largest regions, while the rightmost panel shows their planarity ($P$), filamentarity ($F$) and genus ($G$). }
   \label{Fig.statVAIC}
\end{center}
\end{figure*}

\begin{figure*}
\begin{center}
\includegraphics[scale=0.45]{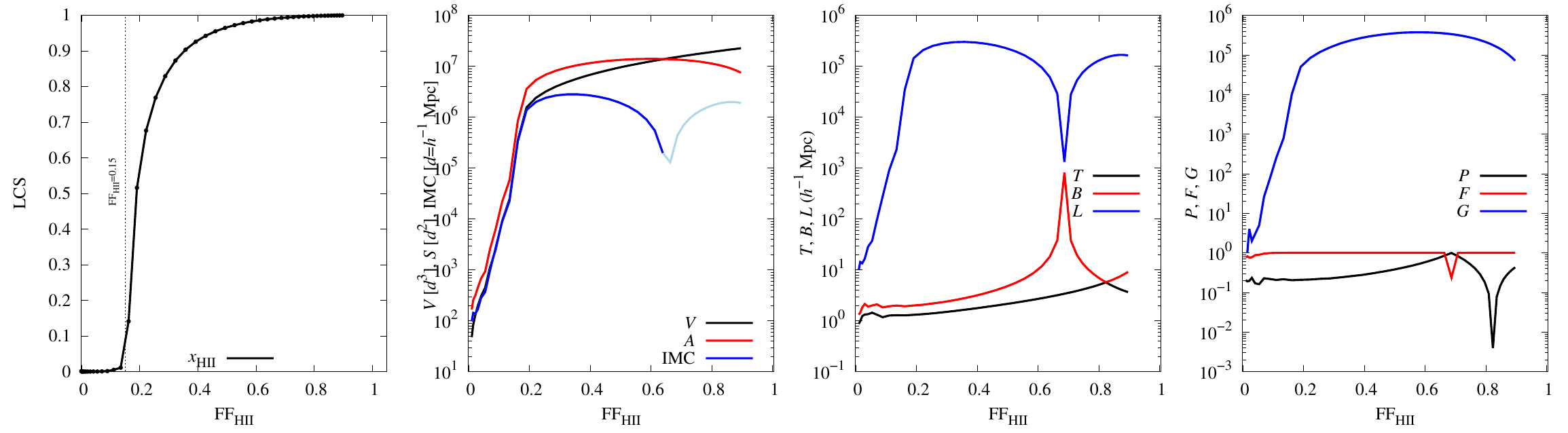}
    \caption{Shape defining quantities corresponding to the ionized regions. The left panel shows the evolution of the Largest cluster statistics (LCS) of the ionized regions as a function of the filling factor of the ionized regions. Evolution of different shape-defining quantities of the largest cluster of ionized regions in the IGM as functions of the filling factor of the ionized regions. The second from the left Panel shows volume ($V$), surface area ($S$) and integrated mean curvature  ($\mathrm{IMC}$). The third from the left panel shows the thickness ($T$), breadth ($B$) and length ($L$) of the largest ionized region. The right panel shows the planarity ($P$), filamentarity ($F$) and genus ($G$) of the largest ionized region. }
   \label{Fig.statVAICXHI}
\end{center}
\end{figure*}

\subsection{Evolution of the largest emission and absorption regions}
\label{sec:lcev}
Next, we study the evolution of shape and morphology of the largest region, separately for the emission and absorption segments, in terms of Shapefinders. Note that we stick to $f_{\mathrm T}=1$ for the rest of the study. This threshold corresponds to the CMB as the radio background. The left panel of Figure \ref{Fig.statVAIC} shows the evolution of the first three Minkowski functionals  $V$, $S$ and $\mathrm{IMC}$ of the largest emission (dashed) and absorption (solid) regions as functions of $\mathrm{FF}_{\rm emi}$ for $f_\mathrm{T}=1$. All these three quantities evolve rapidly around the percolation transition of the corresponding segment (at $\mathrm{FF}_{\rm emi}\approx 0.15$ and $0.95$ for the emission and absorption segments respectively). In fact, this panel demonstrates how the largest emission/absorption region grows rapidly at the onset of their respective percolation transitions and this is manifested in the evolution of the LCS in Figure \ref{Fig.statlcs}.

Let us now focus on the largest emission region. As expected, its $V$ always increases with the increase of $\mathrm{FF}_{\rm emi}$. Its $S$, on the other hand, increases with $\mathrm{FF}_{\rm emi}$ until $\mathrm{FF}_{\rm emi}\sim 0.6$ and slowly decrease thereafter. This shows at stages with $\mathrm{FF}_{\rm emi}\gtrsim 0.6$, the largest emission region in the simulation box becomes less complex in shape with less amount of absorption regions surface inside it, although its volume increases with time.  While $V$ and $S$ are always positive quantities, $\mathrm{IMC}$ can be negative also. The change in the sign of the $\mathrm{IMC}$ occurs around $\mathrm{FF}_{\rm emi}\sim 0.65$ for this model CD.  The negative part of the $\mathrm{IMC}$ is shown in the light-blue colour in the left panel of Figure \ref{Fig.statVAIC}. For $\mathrm{FF}_{\rm emi}\gtrsim 0.7$, one can think of the IGM as a distribution of several absorption regions in an emission region background. At these stages, the sum of $\mathrm{IMC}$s of all the absorption regions is positive which makes the $\mathrm{IMC}$ of the largest emission region negative. 
The evolution of the largest absorption region exhibits similar behaviour but in the opposite direction along the $\mathrm{FF}_{\rm emi}$ axis.

The middle panel of  Figure \ref{Fig.statVAIC} shows the evolution of the three Shapefinders -- `thickness' ($T$), `breadth' ($B$) and `length' ($L$) -- of the largest emission and absorption regions as functions of $\mathrm{FF}_{\rm emi}$, again for $f_\mathrm{T}=1$. The `length' ($L$) of the largest emission or absorption region increases rapidly at the onset of the percolation transition (in the respective segment, i.e. $\mathrm{FF}_{\rm emi}\approx 0.15$ and $0.95$ for the emission and absorption segment respectively) as compared to its $B$ and $T$. The evolution of $L$ of the largest emission region merely follows the evolution of its ${\mathrm{|IMC|}}$ (see Equation \ref{eq.shape}).

The right panel of   Figure \ref{Fig.statVAIC} shows the evolution of morphological quantities, $P$ and $F$, and the topology, $G$, of the largest emission (dashed curves) and absorption (solid curves) regions. $L$ of the largest emission region is more than 2 orders of magnitude larger than its $B$ around the percolation transition stage. This leads to the largest emission region being highly filamentary with $F\sim 1$ (see Equation \ref{eq:PF}) at percolation. The same is also applicable to the largest absorption region.Also, the genus values of both the largest emission and absorption regions increase at the onset of the percolation transition in the respective segments (see the blue curves), and remain stable at post-percolation stages. This implies that the largest regions become more multiply connected with complex topology as they grow and percolate. Note that, at some stage, well after percolation, the genus value of the largest region (emission or absorption) abruptly becomes negative. In this phase where the genus is negative, we show its absolute value by the light-blue coloured curves (solid and dashed). The sign flip, which indicates the change in the topology of the largest region, manifests in negative IMC values (see left panel of Figure \ref{Fig.statVAIC}) which in turn leads to the sharp dip/rise in different Shapefinders (as evident from the middle panel and explained above) at this stage. Indeed, well after percolation in the emission (absorption) segment, many isolated small pockets of absorption (emission) regions emerge inside the percolated largest emission (absorption) region that brings down the latter's genus value and eventually rendering it negative.

\subsection{LCS comparison between the emission and ionized regions}
\label{sec:com}
Next, we compare the LCS of the CD emission regions with the LCS of the EoR ionized regions. We use outputs from the same with {\sc grizzly} simulation to study the LCS of the ionized regions. The ionization history is presented in the left panel of Figure \ref{Fig.tk}.  We consider binary field of ionization fraction $\XHII^{\rm bin}(\mathbfit{x},z)$ between redshifts 20 and 6.5 where $\XHII^{\rm bin}(\mathbfit{x},z)=1$ if $\XHII(\mathbfit{x},z)\geq0.5$. These binary ionization fields are then used as inputs to {\sc surfgen2} for estimating the LCS of the ionized regions. We present the LCS as a function of the filling factor of the ionized regions ($\mathrm{FF}_{\rm HII}$)  in the left panel of Figure \ref{Fig.statVAICXHI}. The percolation of the ionized regions occurs at a stage with  $\overline{x}_{\rm HII}\sim 0.15$ which is consistent with studies such as  \citet[][]{Iliev2006, Chardin2012, Furlanetto2016, 2018MNRAS.477.1984B}. This suggests that the sharp increase of LCS of the emission regions at $\mathrm{FF}_{\rm emi}\sim0.15$ as seen in the left panel of Figure \ref{Fig.statlcs} is similar to the percolation case of the ionized regions.

The second left panel of Figure \ref{Fig.statVAICXHI} shows the evolution of $V$, $S$ and $\mathrm{IMC}$ of the largest ionized region as functions of $\mathrm{FF}_{\rm HII}$. All these three quantities evolve rapidly around $\mathrm{FF}_{\rm HII}\sim0.15$. These are consistent with previous studies such as \citet{2018MNRAS.477.1984B}. Similar to the emission region's LCS case (see the left panel of Figure \ref{Fig.statVAIC}) these quantities evolve sharply around the percolation transition of the ionized segment.  Similar to the sign change for the $\mathrm{IMC}$ of the largest emission region (see the left panel of Figure \ref{Fig.statVAIC}), we also see a sign change for the $\mathrm{IMC}$ of the ionized regions as shown in the second left panel of Figure \ref{Fig.statVAICXHI}. Overall, the evolution of $V$, $S$ and $\mathrm{IMC}$ of the largest emission region around the percolation transition is similar to that of the same quantities of the largest ionized region in the ionization field.

The third left panel of Figure \ref{Fig.statVAICXHI} shows the evolution of $T$, $B$ and $L$ of the largest ionized region as functions of $\mathrm{FF}_{\rm HII}$ while the right panel shows the evolution of $P$, $F$ and $G$. The evolution of $T, ~B$ and $L$ of the largest ionized region in our simulation box as functions of $\mathrm{FF}_{\rm HII}$ is consistent with previous studies such as \citet[][]{2018MNRAS.477.1984B, 2022arXiv220203701P, Dasgupta20233}. These also highlight the filamentary nature of the largest ionized region during ionization percolation and agree well with the evolution of the largest emission region during the CD as we have seen in section \ref{sec:lcev}.


\subsection{Size Distribution of the emission and absorption regions}
\label{BSD-emi}
\begin{figure}
\begin{center}
\includegraphics[scale=0.65]{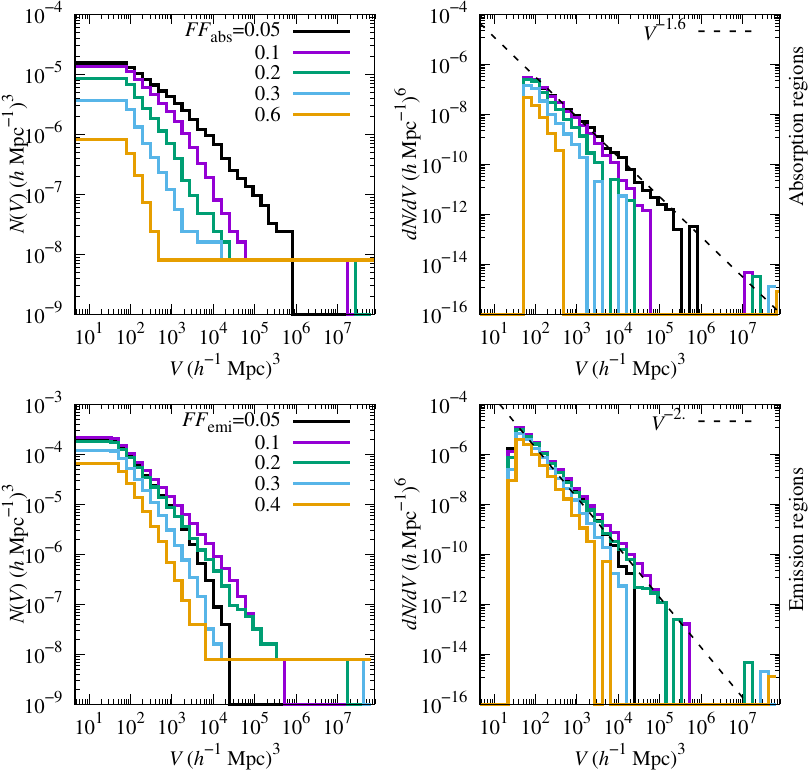}
    \caption{Distribution of emission regions as obtained using {\sc surfgen2} on our simulated Cosmic Dawn scenario. The left panels show the cumulative number of emission (bottom panel) and absorption (top panel) regions as a function of the volume at different stages of the Cosmic Dawn. The right panels show the size distribution of the emission (bottom panel) and absorption (top panel) regions as a function of volume at the same stages of Cosmic Dawn. Different curves in the bottom panels are for $\mathrm{FF}_{\rm emi}=0.05, 0.1, 0.2, 0.3$ and 0.4 which corresponds to redshifts 14.7, 14.3, 13.9, 13.6 and 13.2, respectively. On the other hand, different curves in the top panels are for $\mathrm{FF}_{\rm abs}=0.05, 0.1, 0.2, 0.3$ and 0.6 which corresponds to redshifts 12, 12.3, 12.6, 12.9 and 13.2, respectively. The list of emission and absorption regions is generated from a simulated {\sc grizzly} Cosmic Dawn scenario using {\sc surfgen2} algorithm. }
   \label{fig.bsd}
\end{center}
\end{figure}

The left panels of Figure \ref{fig.bsd} show the cumulative number density $\mathcal{N}(V) \equiv \int_V ^{\infty} \left(\frac{dN}{dV'}\right) dV'$ of absorption (top panel) and emission (bottom panel) regions as functions of volume at different stages of the CD. The right panels show the distribution of the number density $dN/dV$ of the same regions. The individual regions in these distributions are identified in the {\sc surfgen} algorithm using $f_\mathrm{T}=1$. It should be realized that the size distribution of the emission regions is more meaningful when $\mathrm{FF}_{\rm emi}$ is not very large. We show the distribution of the emission regions for  $\mathrm{FF}_{\rm emi} \lesssim 0.5$. For the other half range, i.e.,  $\mathrm{FF}_{\rm emi} \gtrsim 0.5$, we show the distribution of the absorption regions which are more relevant at those stages of the CD.  
 
 Similar to the distribution of the ionized region  \citep[see e.g.,][]{2018MNRAS.477.1984B}, the size distribution of the emission regions also shows a fall of $dN/dV$ towards larger sizes. After the percolation stages, the distribution of the emission regions becomes bimodal. The bimodal nature of $dN/dV$ is also visible for the absorption regions (see the top right panel of Figure \ref{fig.bsd}). As the smaller emission regions overlap and form larger emission regions, the values of $dN/dV$ of the emission regions at the smaller $V$ decrease as the heating progresses. Roughly, $dN/dV$ of the emission regions shows a power-law dependence on the volume, i.e., $dN/dV(V) \propto V^{\tau}$ with $\tau \approx -2$. The same is also true for the absorption regions while $\tau \approx -1.6$ for that. The curves show that a single power-law dependence of $\mathcal{N}(V)$ becomes loose at the lower $V$ side. This shows that the dependencies of the $dN/dV$ on $V$ are not presentable by a single power law and there exit characteristics sizes of the emission (also for the absorption regions) which in this case is $V\sim 20 -30 ~ h^{-1} \rm Mpc$.

There is no unique method to estimate the size distribution of the ionized/emission regions. For example, the friends-of-friends (FOF) algorithm \citep{2006MNRAS.369.1625I} as used in {\sc surfgen2} focuses on the connectivity of a region and provides one large connected emission region well before the end of CD. On the other hand, methods such as the spherical-average \citep{zahn2007}, granulometry algorithm \citep{2017MNRAS.471.1936K} are based on finding the largest spherical volume which fits inside a field and thus provide a very different result compared to FOF. Similarly, algorithms such as the mean-free-path \citep{mesinger07} method which estimates the distribution of the lengths to the edge of a region, find different results. The same statistics also differ for the Watershed \citet{2016MNRAS.461.3361L} algorithm which estimates the effective spherical radius of a region. The definition of bubble size and the interpretation of the size distribution varies with the methodology. A detailed comparison of all these methods in the context of the CD 21-cm signal is beyond the scope of this paper. We refer the readers to \citet{Friedrich2011a} and \citet{2016MNRAS.461.3361L} papers which have compared different methods on EoR ionization maps and pointed out their various advantages and disadvantages. Here, we restrict ourselves to the two well-established methods namely the mean free path (MFP) method \citep{mesinger07} and the granulometry method \citep{2017MNRAS.471.1936K} to obtain the side distributions of the emission and absorption regions and compare the results.

\begin{figure}
\begin{center}
\includegraphics[scale=0.56]{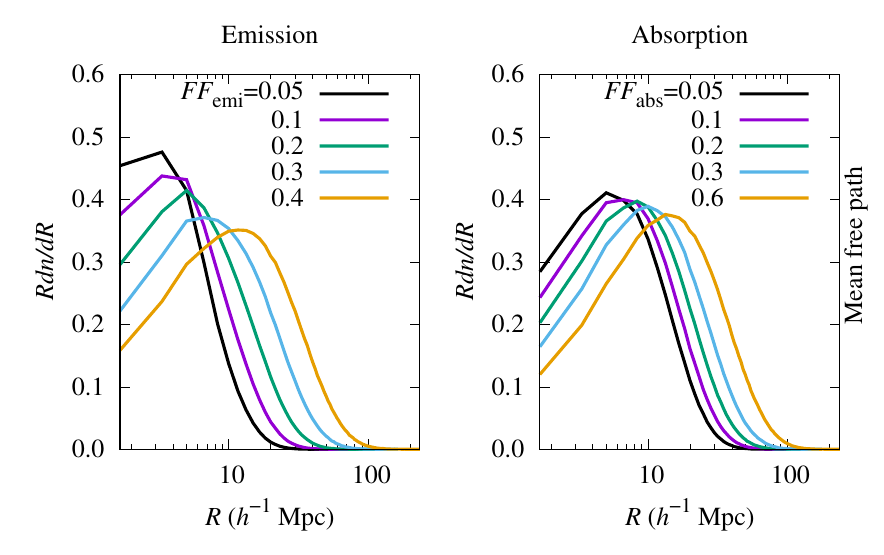}
\includegraphics[scale=0.56]{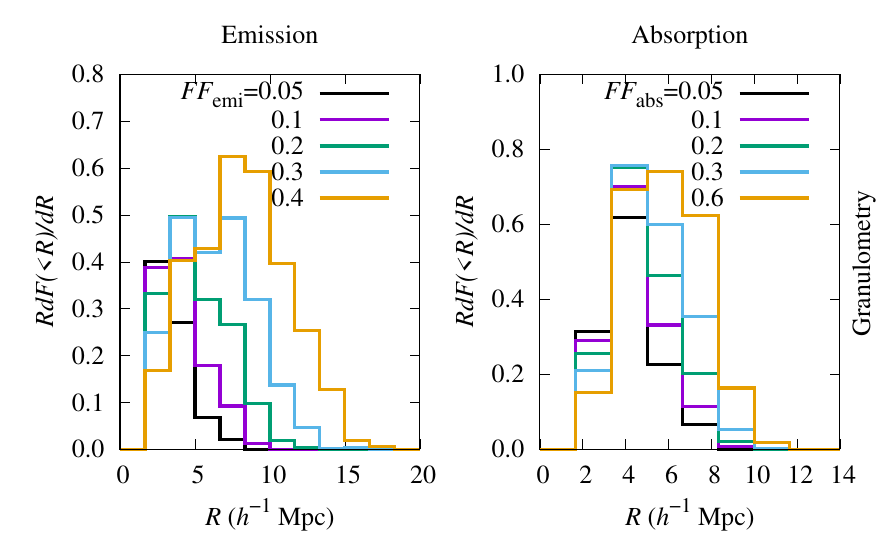}
    \caption{Size distribution of the emission (left panels) and absorption (right panels) regions during the CD. The size distributions in the top panels are estimated using the mean-free path method, while the bottom panels are for the granulometry method. Here, $n$ is the number of rays used in the MFP method whose sizes are between $R$ and $R+dR$. On the other hand, the quantity $F(< R)$ from the granulometry method is the volume fraction of the emission (or absorption) regions whose size is smaller than a radius $R$.}
   \label{image_BSD-MFP-GRAN}
\end{center}
\end{figure}

\begin{figure*}
\begin{center}
\includegraphics[scale=0.5]{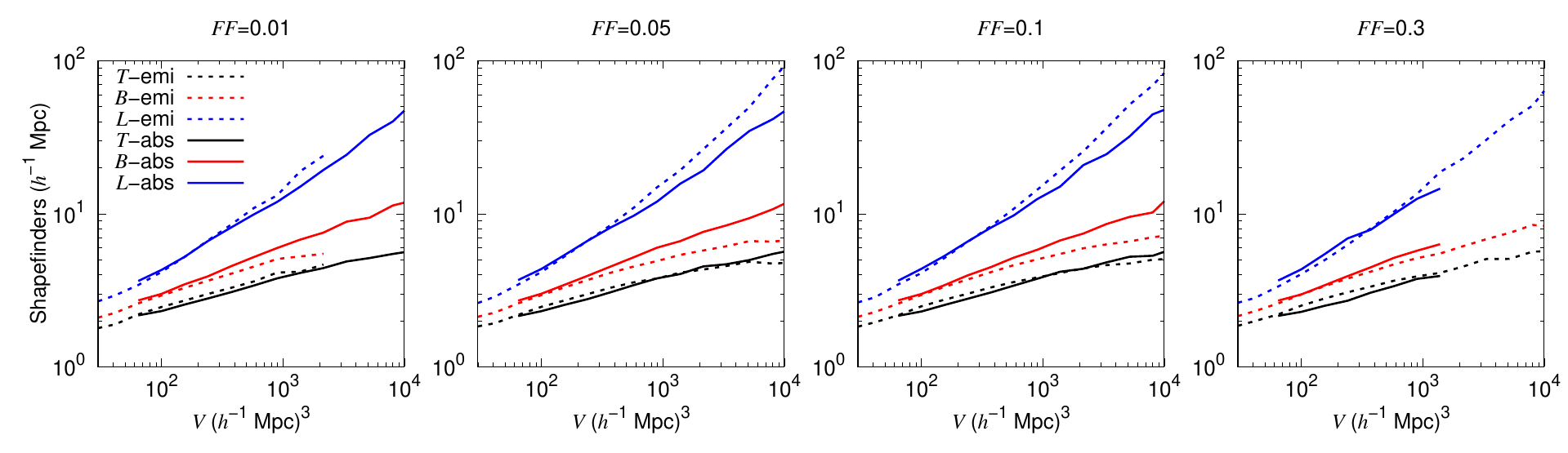}
\includegraphics[scale=0.5]{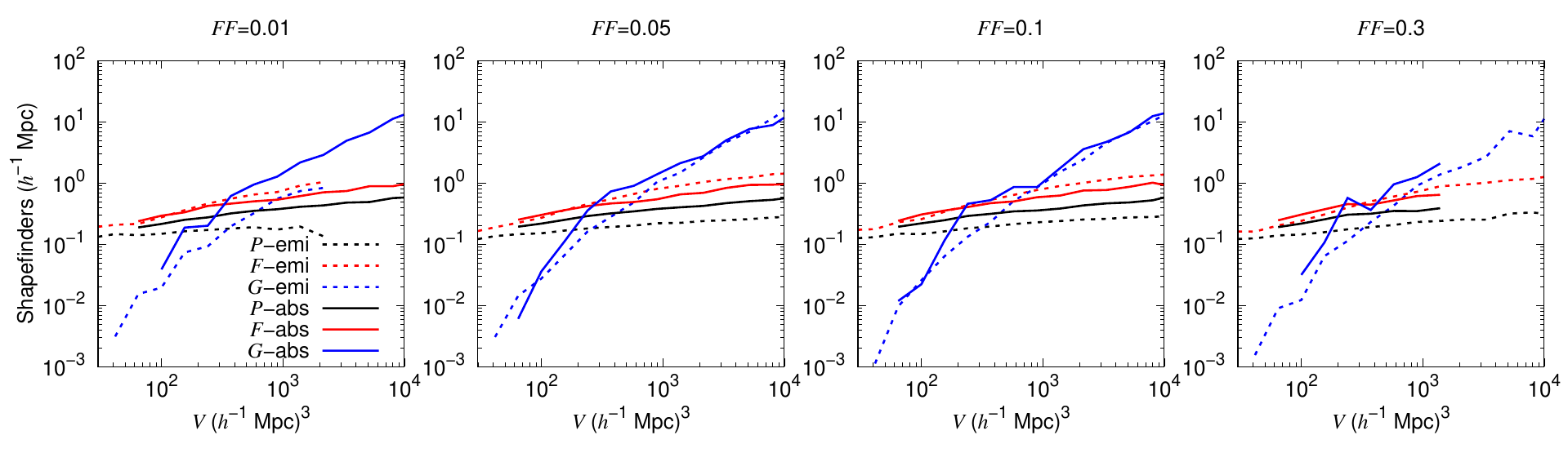}
    \caption{Distribution of the shapefinders at different stages of Cosmic Dawn. The top panels show the Shapefinders `length' ($L$), `breadth' ($B$) and `thickness' ($T$) of the emission (dashed) and the absorption (solid) regions functions of volume ($V$) for the threshold parameter $f_\mathrm{T}=1$ at different stages of the CD. The bottom panels show the `planarity' ($P$), `filamentarity' ($F$) and `genus' ($G$) of the emission and the absorption regions as functions of $V$ for $f_\mathrm{}=1$ at different stages of the CD. $\mathrm{FF}$ stands for the filling factor corresponding to the type of region, i.e. $\mathrm{FF}$ stands for $\mathrm{FF}_{\rm emi}$ ($\mathrm{FF}_{\rm abs}$) when we consider the emission (absorption) regions. }
   \label{Fig:LBTV}
\end{center}
\end{figure*}

For the estimation of the size distribution of these regions using the MFP method, we shoot $10^7$ rays around randomly selected points inside the emission (or absorption) regions in random directions and record the lengths of the rays until those reach the edge of the regions. The top panels of Figure \ref{image_BSD-MFP-GRAN} show the probability density function (PDF) $Rdn/dR$ of the recorded ray lengths at different stages of the CD. Here, $n$ is the number of rays whose sizes are between $R$ and $R+dR$. The top left panel shows the PDFs for the emission regions while the top right panel represents the absorption regions. The overlap of the smaller emission regions at a later stage of heating is prominent by the characteristic scales or the peaks of the PDFs. The peak of the PDFs of the emission regions moves toward the larger values as the heating progresses.  The PDFs show log-normal type behaviour which is similar to the distribution of the ionized regions as well \citep[see e.g.,][]{mesinger07}. The size distribution of the absorption regions also shows a log-normal type nature and is similar to the distribution of the neutral regions during the EoR \citep[see e.g.,][]{2019MNRAS.489.1590G}.

The bottom panels of Figure \ref{image_BSD-MFP-GRAN} show the PDFs $RdF(<R)/dR$ of the size distribution as obtained by using a granulometry method. Here, the quantity $F(< R)$ is the volume fraction of the emission (or absorption) regions whose size is smaller than a radius $R$. We follow the method of \citet{2017MNRAS.471.1936K} which is based on Minkowski subtraction and addition steps to estimate the PDFs. Note that, this method is based on applying spherical filters, and is sensitive to the smallest dimension of a complex-shaped region. Thus, the algorithm does not find any structure with a radius as large as the largest ray length found in the MFP method. In our case, e.g., for $\mathrm{FF}\lesssim 0.6$ the granulometry method does not find any spherical structure of the emission or absorption regions with a radius more than $\sim 20 ~h^{-1} \rm Mpc$.

\subsection{Shape distribution of emission and absorption regions}
In this subsection, we study how the shape (together with topology and morphology) of the emission and absorption regions is distributed over their volume and how these distributions evolve over time. The top panels of Figure \ref{Fig:LBTV} show the distributions of the $T, B$ and $L$ of the emission (dashed curves) and absorption (solid) regions as functions of their volume at different states of the CD. For both types of regions, $L$ is significantly larger compared to their $T$ and $B$ while the difference increases with the increase of volume of the individual regions.  E.g., for $\mathrm{FF}\sim0.1$, $L/T$ or $L/B$ is $\sim 1-2$ at $V\sim 10^2 ~(h^{-1} \rm Mpc)^3$, while the ratios increases to $\sim 20$ for $V\sim 10^4 ~(h^{-1} \rm Mpc)^3$. This implies that small-volume emission/absorption regions are more spherical while the larger regions are more filament-type. The same conclusions can be made from the bottom panels of Figure \ref{Fig:LBTV} as well which show the distribution of $P, F$ and $G$ of the emission (dashed curves) and absorption (solid) regions as functions of their volume for different $\mathrm{FF}$. While $F$ of the emission regions is $\sim 0.2$ for $V\sim 10^2 ~(h^{-1} \rm Mpc)^3$, it increases to $\sim 1$ at $V\sim 2\times10^3 ~(h^{-1} \rm Mpc)^3$. For the same range of $V$, the change of $P$ is smaller compared to that of $F$. From the genus distribution (blue dashed and solid curves) it is evident that larger emission/absorption regions tend to be more multiply connected with more complex topology. Strikingly, the shape (and morphology) distributions of the absorption regions remain similar to that of the emission regions when both segments are compared at the same filling factor.

\section{Summary \& Discussion}
\label{sec:con}
The spatial fluctuation of the redshifted \HI 21-cm signal from 400 million years to the first billion years of our Universe's history carries information about the first radiating sources, the ionization and the thermal states of the IGM. The observations of this signal from the Epoch of Reionization (EoR) and the Cosmic Dawn (CD) using the world's largest radio interferometers have the ability to reveal many unknown facts about these epochs. Theoretical studies suggest that the spatial fluctuations in the signal from the Cosmic Dawn depend on the distribution and properties of the heating sources. An alternative interest focuses on the distribution of the emission regions around the heating sources. Such a morphological study is important to understand the size and distribution of such emission regions or the remaining part of the sky which is seen in absorption. 

In this study, we focus on the shape and morphology of the emission and absorption regions during the Cosmic Dawn. This study is based on a simulation of the Cosmic Dawn 21-cm signal using {\sc grizzly} \citep{ghara15a} code and analyzing it using {\sc surfgen2} algorithm \citep{2019MNRAS.485.2235B} to produce the geometrical properties of the regions. The main findings of our study are the following. 

\begin{itemize}
\item If the radio background is the CMB, the number density of the emission regions gets a maximum at a stage when the volume fraction of regions with gas temperature larger than the CMB temperature (heated fraction $f_{\rm heat}$) is $\sim 0.1$. In case the radio background is stronger than CMB, the peak of the number density vs $f_{\rm heat}$ shifts towards a larger $f_{\rm heat}$ value and can occur between $f_{\rm heat}=0.1$ and $0.9$ depending on the radio background strength. The peak of the number density of the absorption regions as a function of $f_{\rm heat}$ occurs at a stage when $f_{\rm heat}\gtrsim 0.9$.

\item The percolation transition in the emission regions roughly occurs at a filling factor (defined as the volume of the emission segments) of $\mathrm{FF}_{\rm emi} \approx \sim 0.15$. For the absorption regions, the percolation transition occurs at the filling factor $\mathrm{FF}_{\rm abs} \approx 0.05$ which corresponds to $\mathrm{FF}_{\rm emi} \approx \sim 0.95$. The evolution of the `largest cluster statistics' LCS (defined as the ratio of the largest region volume and the total volume of all regions with the same type) of the emission, as well as the absorption regions, as a function of the filling factor, are roughly identical for the different radio background levels chosen in this study. 

\item The overall evolution of the Minkowski functionals -- volume ($V$), surface area ($S$), integrated mean curvature ($\mathrm{IMC}$) and genus ($G$) -- and Shapefinders -- thickness ($T$), breadth ($B$), length ($L$), planarity ($P$) and filamentarity ($F$) -- of the largest emission region (when studied functions of the filling factor of the emission regions) during the CD is found to be similar to the evolution of the same characteristics of the largest ionized region (when studied against ionized filling factor) during the EoR.

\item The number density of emission regions $dN/dV$  as a function of their volume $V$, estimated from the {\sc surfgen2} algorithm, shows a rough power-law dependence on the volume, i.e., $dN/dV(V) \propto V^{\tau}$ with $\tau \approx -2$. We also see similar behaviour for the number density of the absorption regions while $\tau \approx -1.6$ for that. After the percolation transition, the largest region appears as a disjoint part of the size distribution of these regions.

\item The size distribution of the emission regions when estimated using the mean free path method shows a log-normal type behaviour as a function of the length. A similar type of size distribution can be seen for the ionized regions during the EoR. On the other hand, the size distribution of the absorption regions is similar to the size distribution of the neutral regions during the EoR. A similar conclusion can also be made when the size distribution is estimated using the granulometry method.

\item We find that the distributions of the Shapefinders (and genus) of the emission regions, with respect to their volumes, are quite similar to those of the absorption regions when both segments are compared at the same filling factor. In particular, $L$ of both emission and absorption type regions is significantly larger compared to $T$ and $B$.  This difference increases with the increase of the volume of these regions resulting in higher values $F$. Therefore, we find that larger emission/absorption regions tend to be more filamentary with higher values of $G$ as well.

\end{itemize}

The morphology and distribution of both the EoR ionized regions and CD emission regions are crucially dependent on the source properties and the number density of sources during these epochs. For example, a larger value of $M_{\rm min}$ (i.e., the minimum mass of dark matter halos that hosts sources) will result in a smaller number density of X-ray emitting sources and a more patchy heating scenario. The conclusions of the work are based on a single scenario of CD. Thus, it will be important to investigate the robustness of our conclusion for changing CD scenarios. For the reionization case, as shown in \citet{2022arXiv220203701P}, the percolation transition in the ionized regions (as probed by the evolution of LCS of the ionized regions) and the redshift evolution of the Shapefinders can constrain reionization scenarios and shed light on the ionizing sources. Similarly, the redshift evolution of the LCS and Shapefinders in emission regions might distinguish the fundamentally different heating scenarios and put constraints on the properties of the heating sources. We plan to investigate this in a future study.

The simulation volume $\sim (500 ~h^{-1}\rm Mpc)^3$ as used in this work is much larger than the cosmological scale ($\sim 100-200 ~h^{-1} \rm Mpc$) beyond which the cosmological principles of isotropy and homogeneity hold for the EoR and CD 21-cm signal \citep{iliev14, 2020MNRAS.495.2354K}. This ensures that the evolution of the LCS, which detects the percolation transition in the field unambiguously, is independent of the choice of the simulation volume \citep[see][in the context of EoR]{2022arXiv220203701P}. Note that the results of geometrical tools including ours might depend on the resolution of the 21-cm signal fields. \citet{2022arXiv220203701P} investigated the effect of resolution on LCS of ionized regions and found that the LCS results are consistent for different choices of simulation resolutions. We plan to study the effect of simulation resolution on the LCS of the CD 21-cm signal in a follow-up work.

This study has been done in a very idealistic situation. In reality, the 21-cm signal as observed by the radio telescopes will be contaminated by the astrophysical foreground, system noise, radio frequency interference (RFI), etc. Even for the most optimistic scenario with a full $uv$ coverage of the instrument, an efficient method to mitigate the foregrounds, RFI, etc., a perfect calibration technique, the presence of the system noise from a realistic observation (e.g., 1000 hours with the SKA) will affect the recovery of such high-resolution images as we have used in this study. Thus, the recovered size distribution will be affected. Recently, \citet{Dasgupta20233} considered mock SKA1-low observation and studied LCS of EoR 21-cm signal in the presence of telescope noise and other telescope-related effects such as image distortions due to the synthesized beam. The study found that these effects introduce a bias in the redshift evolution of the LCS,  while the percolation transition in the field can be affirmatively recovered even with some calibration error. We expect similar results in the case of CD 21-cm observations as well. Studying the impact of the noise of the system on the recovery of the size distribution of the CD emission regions is beyond the scope of this work. We will address this issue in a follow-up paper.

\section*{Acknowledgements}
SB thanks Varun Sahni, Prakash Sarkar, Santanu Das for their contributions in developing {\sc surfgen2} in its initial phase. RG acknowledges support from the Kaufman Foundation (Gift no. GF01364). SB acknowledges the funding provided by the Alexander von Humboldt Foundation. SM acknowledges financial support through the project titled ``Observing the Cosmic Dawn in Multicolour using Next Generation Telescopes'' funded by the Science and Engineering Research Board (SERB), Department of Science and Technology, Government of India through the Core Research Grant
No. CRG/2021/004025. 

\section*{DATA AVAILABILITY}

The derived data generated in this research will be shared upon reasonable request to the corresponding author.

\bibliography{mybib}

\bsp	
\label{lastpage}
\end{document}